\documentclass{article}
\usepackage{spconf,amsmath,graphicx}
\usepackage{amssymb}
\usepackage{cite}
\usepackage{multirow}
\def\x{{\mathbf x}}														
\def\y{{\mathbf y}}														
\def\W{{\mathbf W}}														
\def\ii{{\hat{\imath}}}												
\def\ij{{\hat{\jmath}}}												
\def\ik{{\hat{\kappa}}}												
\newcommand{\sW}{\mbox{\tiny ${{W}}$}}	  		
\newcommand{\sX}{\mbox{\tiny ${{X}}$}}		 		
\newcommand{\sY}{\mbox{\tiny ${{Y}}$}}				
\newcommand{\sZ}{\mbox{\tiny ${{Z}}$}}				
\DeclareMathOperator{\de}{d} 									
\DeclareMathOperator{\E}{E} 									
\DeclareMathOperator{\var}{var}								
\title{QUATERNION CONVOLUTIONAL NEURAL NETWORKS\\FOR DETECTION AND LOCALIZATION OF 3D SOUND EVENTS}
\name{Danilo Comminiello, Marco Lella, Simone Scardapane, and Aurelio Uncini}
\address{DIET Dept., Sapienza University of Rome\\Via Eudossiana, 18 - 00184 Rome, Italy}
\begin{document}
%
\maketitle
\begin{abstract}
Learning from data in the quaternion domain enables us to exploit internal dependencies of 4D signals and treating them as a single entity. One of the models that perfectly suits with quaternion-valued data processing is represented by 3D acoustic signals in their spherical harmonics decomposition. In this paper, we address the problem of localizing and detecting sound events in the spatial sound field by using quaternion-valued data processing. In particular, we consider the spherical harmonic components of the signals captured by a first-order ambisonic microphone and process them by using a quaternion convolutional neural network. Experimental results show that the proposed approach exploits the correlated nature of the ambisonic signals, thus improving accuracy results in 3D sound event detection and localization.
\end{abstract}
%
%
%
%
%
\section{Introduction}
\label{sec:intro}
Recently, 3D audio processing has been gaining increasing attention due to significant development of spatial audio technology, which paved the way to this emerging field of application. Immersive audio has changed the way people make use of audio services, placing a greater attention to the satisfaction of the user-required quality \cite{EdwardsSPM2018, HuangSPM2011}. In this context, the last few years have been characterized by a wide spread of commercial intelligent acoustic interfaces, basically composed of acoustic interfaces equipped with intelligent signal processors \cite{EdwardsSPM2018, ComminielloAES2013, ComminielloTMM2015}. This kind of devices can be found in many applications, as well as in human everyday life, such as home automation, voice assistance, safety and security by robots, audio surveillance, virtual reality in gaming and entertainment, up to speech recognition applications.

One of the most suited acoustic interfaces for high-definition capturing of the spatial sound field is represented by the Ambisonics, which is basically an array of coincident microphones. The Ambisonics technique is able to capture 3D sounds while minimizing unwanted artifacts caused by cross-talk. 
One of the main features of the Ambisonics is the decomposition of the sound field into a linear combination of spherical harmonics. Traditionally, each ambisonic signal is processed as a separate real-valued signal. However, also due to the physical arrangement of the microphone capsules, ambisonic signals show strongly correlated components. Thus, they lend themselves to a more exotic algebraic description in the quaternion domain that allows signals to be treated as a single multidimensional entity \cite{OrtolaniMLSP2016, OrtolaniUKRCON2017}.

Recently, an increasing interest has been shown on signal processing and machine learning algorithms in quaternion and hypercomplex domains \cite{MizoguchiICASSP2018, XiangTSP2018, OgunfunmiICASSP2018, XiaALMNSM2018, OrtolaniSIGPRO2017, XiaodongSPL2017, SaneiICASSP2018, XiaoICASSP2018}. In such a context, significant advances have been proposed on quaternion neural networks (QNNs) \cite{MinemotoSIGPRO2017, GaudetIJCNN2018, ParcolletARXIV2018, ParcolletIS2018}. In this paper, we want to exploit the capabilities of both QNNs and Ambisonics to analyze 3D sounds, and in particular we focus on the localization and detection of 3D sound events. Both tasks have been widely investigated recently by using convolutional neural networks (CNNs) \cite{ChakrabartyWASPAA2017, FergusonICASSP2018, ThuillierICASSP2018, AdavanneEUSIPCO2018, AdavanneICASSP2017, CakirTASLP2017, JeongDCASE2017}. 
They are also considered as a joint task in \cite{AdavanneARXIV2018} for 3D sounds, but considering each microphone signal as a separate real-valued signal. 

Here, we want to exploit the characteristics of ambisonic signals by processing them as a single multidimensional entity. To this end, we propose a quaternion convolutional neural network (QCNN) for the joint 3D sound event localization and detection (SELD) task. We assess the effectiveness of the proposed method in two different 3D acoustic scenarios and we show improved perfomance for the SELD task with respect to real-valued CNNs proposed in the existing literature.

The paper is organized as follows. In Section~\ref{sec:3dsound}, the representation of the 3D sound field in the quaternion domain is described, while the QCNN is introduced in Section~\ref{sec:qcnn}. Experimental results on SELD problems are shown in Section~\ref{sec:results}. Finally, conclusion are drawn in Section~\ref{sec:conclusion}.

\section{3D Sounds in the Quaternion Domain}
\label{sec:3dsound}
The Ambisonics technique is one of the most popular 3D microphone recording techniques, which is based on a local-space sampling of the sound field by using a coincident microphone array. Such approach involves the decomposition of the sound field into a linear combination of spherical harmonics. 
Here we show how to deal with spherical harmonics and to consider them in the quaternion domain.
%
\subsection{4D Representation of Spatial Sound Fields}
\label{subs:sphericalharmonics}
Spherical harmonics are orthonormal functions which can be used to represent the sound field in terms of its basic components. The sound pressure, in absence of impressed sources, can be expressed by the following wave equation depending on sound speed $c$, radius $r$, azimuth $\theta$ and elevation $\varphi$:
\begin{equation}
	{{\nabla }^{2}}p\left( r,\theta ,\varphi ,t \right)-\frac{1}{{{c}^{2}}}\frac{{{\partial }^{2}}p\left( r,\theta ,\varphi ,t \right)}{\partial {{t}^{2}}}= 0.
	\label{eq:wave_eq}
\end{equation}

\noindent The solution of the wave equation can be achieved by using a Fourier-Bessel series decomposition:
\begin{equation}
	p\left( {\vec r} \right) = \sum\limits_{m = 0}^\infty  {\left(2m+1\right){j^m}{J_m}\left( {kr} \right)\sum_{\substack{0 \le n \le m, \\ \sigma  =  \pm 1}} {X_{mn}^\sigma H_{mn}^\sigma \left( {\theta ,\varphi } \right)  } } 
	\label{Soundfield_Fourier_Bessel_one}
\end{equation} 

\noindent being $m$ the decomposition degree, $n$ the order, $\sigma$ the spin and $k = {2\pi f}/c$ the wave number, ${{J}_{m}}\left( kr \right)$ spherical Bessel functions, and $X_{mn}^\sigma$ a signal component. The previous expression is nothing but a decomposed representation of the sound field. 

Each signal component is weighted by an orthonormal function $H_{mn}^{\sigma }\left( \theta ,\varphi  \right)$, i.e., a \textit{spherical harmonic}, which can be expressed in a normalized form as:
\begin{equation}
	H_{mn}^\sigma \left( {\theta ,\varphi } \right) = {\tilde P_{mn}}\sin\left(\varphi\right)  \times \left\lbrace  \begin{array}{l l} \cos\left(n\theta\right) & \mathrm{{ \: if \:}}\sigma  =  + 1 \\ 
	\sin\left(n\theta\right) & \mathrm{{ \: if \:}}\sigma  =  - 1   \end{array}    \right. .
	\label{eq:sphericalharm}
\end{equation}

\noindent The linear combination of spherical harmonics results in functions on the surface of a sphere. 

Ambisonics is based on the previous description of the sound field by \eqref{Soundfield_Fourier_Bessel_one} and, in particular, it is described by the order $n$, which is also referred to as \textit{ambisonic order}. In this paper, we focus on the so-called \textit{B-Format Ambisonics}, whose order is $n = 1$ (which is the reason why it can be also denoted as first-order ambisonics). The B-Format Ambisonics is composed of an array of 4 coincident microphones (1 omnidirectional and 3 orthogonal figure-of-eight microphones) orthogonal to each other. Each of the 4 microphones is related to a spherical harmonic (1 related to order 0 and 3 to order 1), specifically denoted in this case by $W$ (omnidirectional microphone), $X$, $Y$, $Z$ (figure-of-eight microphones). 
%
%
%
\subsection{Quaternion-Valued Ambisonics Signals}
\label{subs:quaternionambi}
Traditionally, the sound field is defined by spherical harmonics using Euler angles. Here, we aim at dealing with spherical harmonics in the quaternion-valued domain, thus we consider the four ambisonic signals, namely $x_{\sW}\left[n\right]$, $x_{\sX}\left[n\right]$, $x_{\sY}\left[n\right]$ and $x_{\sZ}\left[n\right]$, as a single quaternion signal:
\begin{equation}
	x\left[n\right] = x_{\sW}\left[n\right] + x_{\sX}\left[n\right]\ii + x_{\sY}\left[n\right]\ij + x_{\sZ}\left[n\right]\ik,
	\label{eq:quatinput}
\end{equation}

\noindent which defines a 4-dimensional spatial sound signal, i.e., a quaternion-valued ambisonic signal. In \eqref{eq:quatinput}, the imaginary units, $\ii=\left( 1,0,0 \right)$, $\ij=\left( 0,1,0 \right)$, $\ik=\left( 0,0,1 \right)$, represent an orthonormal basis in ${{\mathbb{R}}^{3}}$ and satisfy the fundamental properties of quaternion algebra \cite{Ward1997}. It is worth noting that, in \eqref{eq:quatinput}, the omnidirectional microphone signal $x_{\sW}\left[n\right]$ is considered as the real component of the quaternion signal, while the three figure-of-eight microphone signals, $x_{\sX}\left[n\right]$, $x_{\sY}\left[n\right]$ and $x_{\sZ}\left[n\right]$ are considered as the imaginary components.

Once defined the expression of the quaternion-valued ambisonic signal, we can use it to effectively process 3D sounds in the quaternion domain, thus fully exploiting the statistical properties of multidimensional signals.
%
%
%
%
%
\section{QUATERNION CONVOLUTIONAL NEURAL NETWORKS FOR 3D SELD}
\label{sec:qcnn}
We introduce now the QCNN method used to jointly perform the 3D SELD task in the quaternion domain when signals are captured by Ambisonics.
%
\subsection{Quaternion-Valued Convolution}
\label{subs:qconv}
The main peculiarity of a QCNN is the convolution process that is performed in the quaternion domain. Here, a quaternion filter matrix is convolved with a quaternion vector by exploiting real-valued representations of quaternions \cite{Ward1997}. Let us consider a quaternion input vector\footnote{We consider monodimensional signals for notational simplicity. As in the real case, everything extends immediately to multidimensional inputs.}, $\x$, defined similarly to \eqref{eq:quatinput}, and a generic quaternion filter matrix defined as $\W = \W_{\sW} + \W_{\sX}\ii + \W_{\sY}\ij + \W_{\sZ}\ik$. The quaternion convolution is obtained from the following Hamilton product:
\begin{equation}
	\begin{split}
		\W \otimes \x &= \left(\W_{\sW}\x_{\sW} - \W_{\sX}\x_{\sX} - \W_{\sY}\x_{\sY} - \W_{\sZ}\x_{\sZ}\right)\\
		&+ \left(\W_{\sW}\x_{\sX} + \W_{\sX}\x_{\sW} + \W_{\sY}\x_{\sZ} - \W_{\sZ}\x_{\sY}\right)\ii \\
		&+ \left(\W_{\sW}\x_{\sY} - \W_{\sX}\x_{\sZ} + \W_{\sY}\x_{\sW} + \W_{\sZ}\x_{\sX}\right)\ij \\
		&+ \left(\W_{\sW}\x_{\sZ} + \W_{\sX}\x_{\sY} - \W_{\sY}\x_{\sX} + \W_{\sZ}\x_{\sW}\right)\ik \\
	\end{split}
	\label{eq:hamprod}
\end{equation}
%
\subsection{Learning in the Quaternion Domain}
\label{subs:qlearn}
The forward phase for a generic quaternion dense layer can be defined by the following expression:
\begin{equation}
	\y = \alpha\left(\W \otimes \x + \mathbf{b}\right)
	\label{eq:qout}
\end{equation}

\noindent where $\y$ is the output of the layer, $\mathbf{b}$ is the quaternion-valued bias offset and $\alpha$ is a quaternion activation function. 
%
The choice of the activation function for the QCNN, as in the real- and complex-valued domains, needs to meet the property of differentiability. A suboptimal but suitable choice is represented by the \textit{quaternion split activation function}, 
defined for a generic quaternion $q$ as:
\begin{equation}
	\alpha\left(q\right) = f\left(q_{\sW}\right) + f\left(q_{\sX}\right) + f\left(q_{\sY}\right) + f\left(q_{\sZ}\right)
	\label{eq:qactfunc}
\end{equation}

\noindent being $f\left(\cdot\right)$ any standard activation function. In our case, we choose $f\left(\cdot\right)$ as a rectified linear unit (ReLU) activation function. The cost function to be optimized is a standard real-valued loss. In particular, in our case, we use a binary class-entropy loss for the SED task and a mean square error (MSE) loss for the localization task, as done in \cite{AdavanneARXIV2018}.
%
\subsection{Weight initialization}
\label{subs:winit}
The appropriate and correct initialization of the network parameters in the quaternion domain must take into account the interactions between quaternion-valued components, thus a simple random and component-wise initialization may result in an unsuitable choice \cite{ParcolletARXIV2018}. Instead, a possible solution may be derived by considering a normalized purely quaternion $u^{\triangleleft}$ generated for each weight $w$ by following a uniform distribution in $\left[0, 1\right]$. Each weight can be written in a polar form as:
\begin{equation}
	w = \left|w\right|e^{u^{\triangleleft} \theta} = \left|w\right|\left(\cos\left(\theta\right) + u^{\triangleleft}\sin\left(\theta\right)\right),
	\label{eq:qweightpolar}
\end{equation}

\noindent from which it is possible to derive the quaternion-valued components of $w$:
\begin{equation}
	\left\{{\begin{array}{*{20}{c}} 
	{{w_{\sW}} = \phi  \cos \left( \theta  \right)}\\
	{{w_{\sX}} = \phi  {u^{\triangleleft}_{\sX}} \sin \left( \theta  \right)}\\
	{{w_{\sY}} = \phi  {u^{\triangleleft}_{\sY}} \sin \left( \theta  \right)}\\
	{{w_{\sZ}} = \phi  {u^{\triangleleft}_{\sW}} \sin \left( \theta  \right)}
	\end{array}} \right.
	\label{eq:qwcomp}
\end{equation}

\noindent where $\theta$ is randomly generated in the range $\left[-\pi, \pi\right]$ and $\phi$ is a randomly generated variable related to the variance of the quaternion weight. The variance of the weight matrix can be defined as $\var\left(\W\right) = \E\left\{\left|\W\right|\right\} - \left(\E\left\{\left|\W\right|\right\}\right)^2$, where the second term is null due to the symmetric distribution of the weight around $0$ \cite{ParcolletARXIV2018}. Since $\W$ follows a Chi distribution with four degrees of freedom, the variance can be expressed as:
\begin{equation}
	\var\left(\W\right) = \E\left\{\left|\W\right|^2\right\} = \int_{0}^{\infty}{w^2 f\left(w\right) \de w} = 4\sigma^2
	\label{eq:qwvar}
\end{equation}

\noindent being $\sigma$ the standard deviation. Denoting with $n_i$ the number of neurons of the input layer and considering the He criterion \cite{HeICCV2015}, $\sigma$ can expressed as $\sigma = {1}/{\sqrt{2n_i}}$ \cite{ParcolletARXIV2018}. 
%
%
It follows that the variable $\phi$ in \eqref{eq:qwcomp} can be randomly generated in the range $\left[-\sigma, \sigma\right]$.
%
\subsection{Network Architecture}
\label{subs:model}
The model receives the quaternion ambisonic input, from which it extracts the spectrogram in terms of magnitude and phase components using a 
Hamming window of length $M$, an overlap of $50\%$, and considering only the $M/2$ positive frequencies without the zeroth bin, similarly to \cite{AdavanneARXIV2018}. Therefore, we obtain a feature sequence of $T$ frames, with an overall dimension of $T \times M/2 \times 8$. The network has a similar architecture to the SELDnet \cite{AdavanneARXIV2018}, in which each input frame is mapped into two parallel outputs, the first one performs the sound event detection (SED), by predicting the active sound event class, and the second one estimates the direction of arrival (DOA) for the detected sound event by a multi-class regression. 

In particular, each input frame is processed by the neural network in which the learning of the local shift-invariant features of the spectrogram is performed by using multiple layers of 2D QCNN. The QCNN layers are composed of $P$ filter kernels with size $3 \times 3 \times 8$ and ReLU activation functions. 
At the output of the activation function a batch normalization is performed and a max-pooling is applied along the frequency axis for dimensionality reduction while preserving the sequence length $T$. The output of the final QCNN layer has a dimension of $T \times 2 \time 4P$, where the frequency dimension $2$ iss reduced by the max-pooling, while the number of output feature maps is $4$ times larger, with respect to a standard CNN, due to the quaternion convolution. The output of the QCNN is reshaped into a $T \times 8P$ frame, which is then processed by a bidirectional recurrent neural network, as in the SELDnet, with the aim of learning the temporal information. 
Then, two branches of fully connected layers are used in parallel, one for each task. 
The first layer in both the branches involves $R$ nodes with linear activation functions, while the last layer for the branch related to the SED task has $N$ nodes, each one corresponding to a sound event class to be detected. A sigmoid function is used for multi-class detection, i.e., multiple sounds detected simultaneously. On the other hand, the last layer of the branch related to the localization task involves $3N$ nodes, representing the Cardinal coordinates for each sound event class, and hyperbolic tangent activation functions. 
As for the SELDnet, we use a cross-validation for the hyperparameter optimization. The network training involves a weighted combination of binary cross-entropy and MSE using Adam optimizer as also done in \cite{AdavanneARXIV2018}. 
%
%
%
%
%
\section{EXPERIMENTAL RESULTS}
\label{sec:results}
%
\subsection{Datasets}
\label{subs:dataset}
We evaluate the proposed method involving the QCNN on two datasets involving 3D sound events in the Ambisonics format recorded in anechoic and reverberant environments. Both the datasets consider stationary sources associated with spatial coordinates. 

The first dataset is the \textit{Ambisonic, Anechoic and Synthetic Impulse Response} (ANSYN) dataset \cite{AdavanneARXIV2018, AdavanneEUSIPCO2018}, consisting of spatially located sound events in an anechoic scenario using simulated impulse responses. The dataset is divided in three subsets, O1, O2, O3, involving respectively a maximum number of 1, 2 and 3 simultaneously active sound events. Each subset is composed of three validation splits with 240 training and 60 testing Ambisonics recordings, each one during 30 seconds at 44100 Hz. The dataset contains 11 isolated sound event classes, each one composed of 20 examples, 16 of which randomly chosen for the training set and the remaining 4 are used for the test set. 

The second dataset is the \textit{Ambisonic, Reverberant and Synthetic Impulse Response} (RESYN) dataset, similar to the ANSYN with the only difference that the environment is reverberant. Indeed, a room of size $10 \times 8 \times 4$ m is considered with reverberation times 1.0, 0.8, 0.7, 0.6, 0.5 and 0.4 for each octave band, and 125 to 4000 Hz band center frequencies. More details on the datasets can be found in \cite{AdavanneEUSIPCO2018}.
%
\subsection{Metrics}
\label{subs:metrics}
The SELD task can use individual SED and localization metrics \cite{AdavanneARXIV2018}. For the SED task, we use the polyphonic SED metrics that are the F-score (ideally $F = 1$), based on the number of true and false positives, and the error rate (ER) (ideally $ER = 0$), based on the total number of active sound event classes in the ground truth. A joint SED score can be considered as $S_{\text{SED}} = \left(ER + \left(1 - F\right)\right)/2$.

On the other hand, a DOA estimation error $DOA_{\text{err}}$ can be used as evaluation metric for localization task, based on estimated and ground truth DOAs \cite{AdavanneARXIV2018}. Moreover, a frame recall metric $K$ (ideally $K = 1$) can be used based on the percentage of true positives. A joint DOA score can be defined as $S_{\text{DOA}} = \left(DOA_{\text{err}}/180 + \left(1 - K\right)\right)/2$.

Finally, an overall SELD score can be defined based on the previous metrics as $S_{\text{SELD}} = \left(S_{\text{SED}} + S_{\text{DOA}}\right)/2$.
%
\subsection{Evaluation}
\label{subs:evaluation}
We compare the proposed quaternion model with the SELDnet architecture \cite{AdavanneARXIV2018} on the ANSYN and RESYN datasets. In order to provide a fair comparison, we use a configuration such to have a comparable number of parameters for both the models. In particular, we have about 760k parameters for the proposed quaternion network and about 530k parameters for the SELDnet, which is the most similar configuration possible considering the higher number of parameters generated by the QCNN, as described in Section~\ref{sec:qcnn}. To this end, we set a number of $P = 64$ filters, sequence length of $T = 512$ frames, window length $M = 512$, batch size of $16$, $Q = 128$ nodes for the recurrent networks and $R = 32$ nodes for the fully connected layers. The models have been trained over 1000 epochs\footnote{Experiments were run thanks to TensorFlow Research Cloud.}.
%

Results for the ANSYN dataset are shown in Table~\ref{table:ansyn}. In terms of the overall SELD score, the proposed quaternion method clearly outperforms the standard SELDnet in each validation split and considering different overlapping sounds. In particular, it is worth noting from Table~\ref{table:ansyn} that, while achieving better performance also in terms of localization score, the most significant part of the improvement is represented by the SED score, which is largely reduced with respect to the standard SELDnet. 
Similar conclusions can be drawn also from the results achieved for the RESYN dataset and shown in Table~\ref{table:resyn}. It can be noted that scores are slightly worse with respect to previous results due to reverberations. However, even in this case, the proposed quaternion method is able to improve both individual SED and DOA scores and the overall SELD performance.  
\begin{table}[t]
	\caption{Results on the ANSYN dataset in terms of the SED, DOA and overall SELD score. Best SELD scores in bold.}
	\label{table:ansyn}
	\vspace{0.3cm}
	\begin{tabular}{cc|ccc|ccc}
		\multicolumn{2}{c|}{}               			& \multicolumn{3}{c|}{SELDnet} & \multicolumn{3}{c}{Proposed Method}          \\ \hline
		\multicolumn{2}{c|}{Val. split} 					& 1        & 2       & 3       & 1             & 2             & 3             \\ \hline
		\multirow{3}{*}{O1} & $S_{\text{SED}}$  	& 0.22     & 0.21    & 0.31    & 0.14          & 0.12          & 0.16          \\
												& $S_{\text{DOA}}$    & 0.20     & 0.21    & 0.21    & 0.12          & 0.10          & 0.10          \\
												& $S_{\text{SELD}}$ 	& 0.21     & 0.21    & 0.26    & \textbf{0.13} & \textbf{0.11} & \textbf{0.13} \\ \hline
		\multirow{3}{*}{O2} & $S_{\text{SED}}$    & 0.47     & 0.44    & 0.47    & 0.33          & 0.33          & 0.34          \\
												& $S_{\text{DOA}}$    & 0.35     & 0.34    & 0.33    & 0.29          & 0.29          & 0.30          \\
												& $S_{\text{SELD}}$ 	& 0.41     & 0.39    & 0.40    & \textbf{0.31} & \textbf{0.31} & \textbf{0.32} \\ \hline
		\multirow{3}{*}{O3} & $S_{\text{SED}}$    & 0.53     & 0.57    & 0.55    & 0.48          & 0.46          & 0.45          \\
												& $S_{\text{DOA}}$    & 0.47     & 0.45    & 0.45    & 0.42          & 0.40          & 0.41          \\
												& $S_{\text{SELD}}$ 	& 0.50     & 0.51    & 0.50    & \textbf{0.45} & \textbf{0.43} & \textbf{0.43}
	\end{tabular}
\end{table}

%
\vspace{0.3cm}
\begin{table}[t]
	\caption{Results on the RESYN dataset in terms of the SED, DOA and overall SELD score. Best SELD scores in bold.}
	\label{table:resyn}
	\vspace{0.3cm}
	\begin{tabular}{cc|ccc|ccc}
		\multicolumn{2}{c|}{}               			& \multicolumn{3}{c|}{SELDnet} & \multicolumn{3}{c}{Proposed Method}          \\ \hline
		\multicolumn{2}{c|}{Val. split} 					& 1        & 2       & 3       & 1             & 2             & 3             \\ \hline
		\multirow{3}{*}{O1} & $S_{\text{SED}}$  	& 0.22     & 0.24    & 0.30    & 0.23          & 0.22          & 0.29          \\
												& $S_{\text{DOA}}$    & 0.38     & 0.24    & 0.26    & 0.27          & 0.24          & 0.26          \\
												& $S_{\text{SELD}}$ 	& 0.30     & 0.24    & 0.28    & \textbf{0.25} & \textbf{0.23} & \textbf{0.27} \\ \hline
		\multirow{3}{*}{O2} & $S_{\text{SED}}$    & 0.57     & 0.54    & 0.61    & 0.47          & 0.40          & 0.46          \\
												& $S_{\text{DOA}}$    & 0.45     & 0.46    & 0.41    & 0.43          & 0.41          & 0.42          \\
												& $S_{\text{SELD}}$ 	& 0.51     & 0.50    & 0.51    & \textbf{0.45} & \textbf{0.41} & \textbf{0.44} \\ \hline
		\multirow{3}{*}{O3} & $S_{\text{SED}}$    & 0.64     & 0.59    & 0.57    & 0.51          & 0.53          & 0.55          \\
												& $S_{\text{DOA}}$    & 0.46     & 0.49    & 0.49    & 0.47          & 0.49          & 0.51          \\
												& $S_{\text{SELD}}$ 	& 0.55     & 0.54    & 0.53    & \textbf{0.49} & \textbf{0.51} & \textbf{0.53}
	\end{tabular}
\end{table}
%
%
%
%
%
\vspace{-1em}
\section{CONCLUSION}
\label{sec:conclusion}
In this paper we propose a SELDnet method involving a QCNN for the detection and the localization of 3D sound events captured by first-order Ambisonics. Ambisonic microphone signals are represented in their spherical harmonics form, which enables the processing in the quaternion domain. In particular, the convolution process of the neural network is performed in the quaternion domain, as well as the learning. Results are evaluated on the ANSYN and RESYN datasets and they have shown that, due to the processing in the quaternion domain, the proposed method is able to exploit the correlated nature of the ambisonic signals, thus providing improvements with respect to the standard SELDnet in terms of the simultaneous detection and localization scores.

%
%
%
%
%
%
%
%
%
%
%
\ninept
\bibliographystyle{IEEEbib}
\bibliography{ICASSP19refs}
%
%
\end{document}